\documentclass[12pt,tightenlines,eqsecnum,floats,aps,amsmath,amssymb,nofootinbib,prd,showpacs]{revtex4}

\usepackage{amsmath,amssymb,amsfonts}
\usepackage{graphicx}
\usepackage{enumerate} 
\usepackage{colordvi} 

\def\be{\begin{equation}}
\def\ee{\end{equation}}
\def\ba{\begin{eqnarray}}
\def\ea{\end{eqnarray}}

\def\R{\mathbb{R}}

\def\k{\kappa}
\def\M{\mathcal{M}}
\def\H{\mathcal{H}}
\def\L{\mathcal{L}}
\def\ez{\,{}^o\! e}
\def\dd{\textrm{d}}

\def\b{\bar}
\def\f{\frac}
\def\p{\partial}
\def\h{\hat}

\def\t{\tilde}
\def\wt{\widetilde}

\newcommand{\D}[0]{\rlap{$\mathrm{d}$}\hspace{0.3ex}\mathrm{d}}

\usepackage{enumerate}

\usepackage{colordvi}
\newcounter{mnotecount}[section]

\newcommand{\comment}[1]{}

\begin{document}
\preprint{\vbox{\baselineskip=12pt \rightline{IGC-08/01-}
}}
\title{Asymptotics and Hamiltonians in a First order formalism}

\author{Abhay Ashtekar${}^1$}\email{ashtekar@gravity.psu.edu}
\author{Jonathan Engle${}^{1,2}$}\email{engle@cpt.univ-mrs.fr}
\author{David Sloan${}^1$}\email{sloan@gravity.psu.edu}
\affiliation{${}^{1}$Institute for Gravitation and the Cosmos, Penn State,
University Park, PA 16802, U.S.A. \\ ${}^2$ Albert Einstein Institute,
Am M\"ulenberg 1, 14776 Golm, Germany}

\begin{abstract}

We consider 4-dimensional space-times which are asymptotically
flat at spatial infinity and show that, in the first order
framework, action principle for general relativity is well-defined
\emph{without the need of infinite counter terms.} It naturally
leads to a covariant phase space in which the Hamiltonians
generating asymptotic symmetries provide the total energy-momentum
and angular momentum of the space-time. We address the subtle but
important problems that arise because of logarithmic translations
and super-translations both in the Langrangian and Hamiltonian
frameworks. As a forthcoming paper will show, the treatment of
higher dimensions is considerably simpler. Our first order
framework also suggests a new direction for generalizing the
spectral action of non-commutative geometry.

\end{abstract}


\maketitle

\section{Introduction}
\label{s1}

In most field theories the action depends only on fundamental
fields and their first derivatives. By contrast, the
Einstein-Hilbert action of general relativity depends also on the
second derivative of the fundamental field, the space-time metric
$g$. As a consequence, stationary points of this action do not
yield Einstein's equations unless both the metric and its first
derivatives are kept fixed at the boundary; strictly we do not
have a well-defined variational principle. To remedy this
situation, Gibbons and Hawking \cite{gh,swh} proposed that we add
a surface term to the Einstein Hilbert action. We are then led to
\be \label{gh} S_{\rm EH+GH} (g)\, = \,  \f{1}{2\k} \left( \int_{\M}
R\, d^4V \,+\, 2\int_{\p\M} K \, d^3V + C \right)\, .\ee
Here $\k = 8\pi G$,\, $\M$ is a 4-manifold representing an
appropriate portion of space-time, $\p\M$ its boundary, $R$ the
Ricci scalar of the metric $g$,\, $K$ the trace of the extrinsic
curvature of $\p\M$, and $C$ is an arbitrary function of the
metric $h$ induced on $\p\M$ by $g$.

Let us restrict ourselves to cases where $g$ has signature -,+,+,+
and is smooth and globally hyperbolic. We will let $\M$ be the
space-time region bounded between two Cauchy surfaces. If $\M$ is
spatially compact, by setting $C=0$, we obtain a well-defined
variational principle and a finite on-shell action. However, in the
asymptotically flat case, it is well-known that this strategy has
some important limitations (see e.g., \cite{mm}). In particular, the
action is typically infinite even `on-shell', and indeed even when
$g$ is the Minkowski metric. To remedy this problem, Gibbons and
Hawking \cite{gh,swh} proposed an infinite subtraction: Carry out an
isometric embedding of $(\p\M, h)$ in Minkowski space, calculate the
trace $K_o$ of the extrinsic curvature of $\p\M$ defined by the
Minkowski metric, and set $C = -K_o$. Can this procedure be carried
out for generic asymptotically flat Lorentzian metrics $g$? Now,
there is a theorem due to Weyl which guarantees that any 2-manifold
with a metric of positive scalar curvature can be isometrically
embedded in the Euclidean 3-space. One may imagine that this result
could be extended to higher dimensional Minkowski spaces. However,
this expectation is not borne out. For, in a $d$ dimensional
space-time the metric $h$ on the boundary $\p\M$ has $d(d-1)/2-
(d-1)$ degrees of freedom (after removing the diffeomorphism gauge)
while the choice of embedding provides a freedom worth only one
function on $\p\M$. Thus, even at this heuristic level, if $d \ge 4$
the freedom is not sufficient whence this infinite subtraction
procedure will not work for generic metrics $g$.

Over the last few years, a new set of proposals for infinite
counter terms $C$ have appeared in the literature, motivated in
part by earlier work of Brown and York \cite{by}. In particular,
Kraus, Larsen, Siebelink \cite{kls} have constructed a
counter-term using a (non-polynomial) function of the Ricci
curvature of the boundary. Mann and Marolf \cite{mm} have
introduced a counter-term which is closer to the spirit of the
Gibbons-Hawking proposal. They replace $K_o$ with the trace of a
tensor field $\h{K}_{ab}$ which generalizes the extrinsic
curvature $K_{ab}^o$ of $\p\M$ with respect to the Minkowski
metric, used by Gibbons and Hawking, to situations in which the
boundary may not be isometrically embedable in Minkowski space.
The Mann-Marolf procedure is motivated by the form of the Gauss
Codazzi equations and is carried out entirely in the physical
space-time, without recourse to any embedding. Not only do these
improved actions $S_{\rm imp}$ lead to well-defined action
principles, but they also overcome another limitation of the
original proposal: Now $\delta S_{\rm imp} =0$ at asymptotically
flat solutions for \emph{all} permissible variations $\delta$.

Since we are dealing just with \emph{classical} field theories where
all fields are smooth, one might wonder if there is a way to avoid
infinite subtractions altogether and construct an action principle
which is manifestly finite from the beginning. The first goal of
this paper is to show in some detail that this is indeed possible if
one uses a first order framework based on orthonormal tetrads and
Lorentz connections.%
\footnote{This fact was used in \cite{afk,abl} where, however,
attention was focused on the inner, isolated horizon boundary
rather than at infinity.}
It is worth noting that this framework constitutes the starting
point of Hamiltonian loop quantum gravity as well as spin-foams. The
second goal of the paper is to use this action to construct a
covariant Hamiltonian framework by keeping careful track of boundary
conditions.%
\footnote{For a complementary treatment in the canonical
framework, see \cite{aa87,tt}. However, that analysis is based on
self-dual, rather than real Lorentz-connections and did not
address issues involving the action principle discussed above.}
It is well-known that the asymptotic structure is quite subtle in
4-dimensions because of the presence of logarithmic translations
and super-translations \cite{ah,aa-log}. These subtleties lead to
considerable complications in the definition of asymptotic
symmetries and conserved quantities. The ensuing difficulties were
overcome in discussions of conserved quantities based on field
equations quite sometime ago \cite{ah,bs,aa-log,aa-ein,ar}.
However, in much of the older literature on the derivation of
Hamiltonians from action, these subtleties were generally
overlooked (see, e.g., \cite{adm,titelboim,hh}). We will fill this
gap starting with the first order action principle, thus providing
a self-contained Lagrangian and Hamiltonian description without
the need of any infinite counter-terms. The first order framework
also has the technical advantage that calculations can be carried
out using exterior calculus and tend to be significantly simpler.

The paper is organized as follows. In section \ref{s2} we recall
the asymptotic structure at spatial infinity in 4 dimensions and
in section \ref{s3} we introduce the Lagrangian and Hamiltonian
framework using the first order framework. A careful choice of
boundary conditions eliminates the logarithmic and super
translations and reduces the asymptotic symmetry group to
Poincar\'e. In section \ref{s4} we calculate the Hamiltonians
generating these Poincar\'e symmetries and in section \ref{s5} we
discuss their properties. In particular, although they are now
defined using entirely different techniques these expressions of
Poincar\'e-momentum agree with the earlier ones
\cite{ah,bs,aa-ein,ar} obtained using field equations. In this
respect results of section \ref{s5} are similar to those obtained
by Mann, Marolf and Virmani \cite{mm,mmv} using their infinite
counter-term subtraction in the second order framework. Section
\ref{s6} summarizes the results and discusses their relevance to
recent developments concerning the spectral action used in
non-commutative geometry \cite{cc1,cc2}.

\section{Asymptotic Structure:\, Subtleties in 4 dimensions}
\label{s2}

In this section we will introduce some notation, specify our
boundary conditions at spatial infinity and discuss some important
complications that are peculiar to 4 dimensions.

Let $\eta$ be a Minkowski metric on $\R^4$. Since we are primarily
interested in spatial infinity, we will focus on the region
$\mathcal{R}$ which lies outside the light cone of some point $p$
in the interior. On $\mathcal{R}$ the 4-dimensional radial
coordinate $\rho$ is given by $\rho^2 = \eta_{ab} x^a x^b$ where
$x^a$ are the Cartesian coordinates of $\eta$ with $p$ as origin.
Functions $f$ of interest will admit a power series expansion of
the type
\be \label{falloff} f(\rho,\Phi)= \sum_{n=0}^{m}\, \f{{}^{n}\!
f(\Phi)}{\rho^{n}}\, +\, o(\rho^{-m})\ee
where $\Phi = (\theta,\phi, \chi)$ are the standard angles on
hyperboloids $\H$ defined by $\rho = {\rm const}$ and where the
remainder $o(\rho^{-m})$ has the property that $\lim_{\rho
\rightarrow \infty}\,\, \rho^m\,o(\rho^{-m}) =0$. Such a function
will be said to \emph{admit an asymptotic expansion to order $m$}.
A tensor field $T^{a...b}{}_{c...d}$ will be said to admit an
asymptotic expansion to order $m$ if all its components in the
\emph{Cartesian} chart $x^a$ do so. Derivatives\, $\partial
T^{a...b}{}_{c...d}/\partial x^e$\, of such a tensor field with
respect to $x^e$ will be assumed to admit an asymptotic expansion
to order $m+1$. Finally, note that the remainder in
(\ref{falloff}) can contain terms of the type $(h(\Phi)\, \ln
\rho)/\rho^{m+1}$ whence $ \lim_{\rho \rightarrow
\infty}\,\rho^{m+1} o(\rho^{-n})$ need not exist.

A smooth space-time metric $g$ on $\mathcal{R}$ will be said to be
\emph{weakly asymptotically flat at spatial infinity} if there
exists a Minkowski metric $\eta$ such that outside a spatially
compact world tube $g-\eta$ admits an asymptotic expansion to order
1 and $\lim_{\rho\rightarrow\infty}\, (g-\eta) =0$. This condition
implies that in the $(\rho,\Phi)$ chart associated with $\eta$,\,
the line element of $g$ can be expanded as follows:
\be \label{g1} g_{ab} \dd x^a \dd x^b  = \left(1+
\f{2\sigma(\Phi)}{\rho}\right)\, \dd\rho^2 \, +\, 2\rho\,
\f{A_i(\Phi)}{\rho}\,\, \dd\rho\, \dd\Phi^i\, +\, \rho^2\,
\left(h_{ij} + \f{{}^1\!h_{ij}}{\rho}\right)\, \dd\Phi^i \dd\Phi^j
\,+\, o(\rho^{-1})\, \ee
where $h_{ij}$ is the metric on the unit (time-like) hyperboloid
in Minkowski space. Given such an expansion, it is tempting to
think of $\eta$ as a reference metric, consider its Poincar\'e
group as the asymptotic symmetry group of $g$ and define
Poincar\'e momentum ---i.e., energy-momentum and relativistic
angular momentum--- using Hamiltonians generating these
transformations on the appropriate phase space. Indeed, this is
often done in the literature. However, the problem is that the
metric $\eta$ is not uniquely singled out by the physical metric
$g$. For, if $g$ admits an asymptotic expansion (\ref{g1}) with
respect to one Minkowski metric $\eta$, it also admits this
expansion with respect to another Minkowski metric $\bar\eta$ if
$\eta-\bar\eta$ admits an asymptotic expansion to order $1$ and
$\lim_{\rho\rightarrow\infty} (\eta-\b\eta) =0$. Such Minkowskian
metrics will be said to be \emph{compatible}. One might first
think that the Poincar\'e groups of compatible metric would agree
asymptotically. Unfortunately, as we now show, this is not the
case in 4 dimensions.

Set
\be \bar{x}^a = L_a{}^b x^b + \ln \rho\, C^a  + T^a + S^a(\Phi) +
o(\rho^0) \ee
where $L_a{}^b$ is a Lorentz transformation (i.e. $L_a{}^b\,
L_c{}^d \eta_{bd} = \eta_{ac}$), and $T^a, C^a$ are constant
vector fields with respect to $\eta$. Let $\b{\eta}^{ab}$ be the
Minkowski metric for which $\b{x}^a$ is the Cartesian chart. Then
it is easy to check that $\eta$ and $\b\eta$ are compatible. If
$C^a $ and $S^a(\Phi)$ were set to zero, $\bar{x}^a$ would be just
another Cartesian chart for $\eta$ to leading order whence
Poincar\'e groups of $\bar\eta$ and $\eta$ would have agreed
asymptotically. The term $C^a \ln \rho$ defines a
\emph{logarithmic translation} while the angle dependent
translation $S^a(\phi)$ is called a \emph{super-translation.} If
$C^a$ and/or $S^a(\Phi)$ are non-zero, the Poincar\'e groups of
$\eta$ and $\b\eta$ fail to agree even at infinity. Since one can
define Poincar\'e momentum using any of them, there is
considerable ambiguity in the values of conserved quantities. The
ambiguity introduced by logarithmic translations was first pointed
out by Bergmann \cite{pgb} while that associated with
super-translations was first observed at null infinity by Bondi
and Sachs \cite{bms} and later at spatial infinity in \cite{ah}.
These complications arise because, for physical reasons, we can
only ask that $g-\eta$ should fall off as $(\rho^{-1})$ in 4
dimensions. Had we demanded that they fall off as $(\rho^{-2})$,
we would have been forced to set $C^a$ and $S^a(\Phi)$ to zero and
the ambiguities would have disappeared. But of course, because of
the positive energy theorem, in this case the only solution
satisfying our putative boundary conditions would have been
Minkowski space! In $d$ space-time dimensions, the physically
correct condition is that $g-\eta$ should fall off as
$(\rho^{-(d-3)})$, whence there is no freedom to perform
super-translations or logarithmic translations if $d>4$.

The challenge in 4-dimensions, then, is to strengthen the boundary
conditions so that this freedom is eliminated without excessively
limiting permissible space-time geometries. Let us begin with the
logarithmic translations. It is straightforward to verify that under
$x^a \rightarrow \bar{x}^a = \ln \rho\, C^a$, we are led to a
Minkowski metric $\b\eta$ and hence the `barred'-version of the
expansion (\ref{g1}) in which
\be \b{\sigma}(\Phi) = \sigma(\Phi) + C^a \h{x}_a \quad {\rm
where} \quad \h{x}_a = \f{\eta_{ab}\, x^b}{\rho}\, .\ee
Note that for any non-zero $C^a$,\, $C^a \h{x}_a$ is function on the
hyperboloids $\H$ which is \emph{odd under the reflection} $\h{x}^a
\rightarrow -\h{x}^a$. Therefore, one can eliminate the freedom to
perform the logarithmic translations \emph{by simply demanding that
in the asymptotic expansion (\ref{g1}), $\sigma(\Phi)$ be even under
these reflections} \cite{aa-log}.  How stringent is this requirement
on $\sigma$? Analysis of the asymptotic structure at spatial
infinity at $i^o$ ties it to the asymptotic behavior of the
`electric' part $E_{ab} = C_{acbd} \h{x}^c \h{x}^d$ of the Weyl
tensor. The leading asymptotic part of $E_{ab}$ is a tensor field
tangential to $\H$ given by
\be {}^3\!E_{ab} (\Phi):= \lim_{\rho \rightarrow \infty}\, \rho^3
E_{ab}\, .\ee
It admits a scalar potential which is precisely $\sigma$:\,
${}^3\!E_{ab} = -(D_a D_b \sigma + \sigma h_{ab})$, where $D$ is
the derivative operator of $(\H, h_{ab})$ \cite{aa-ein,bs,ar}. Now
one can show that if ${}^3\!E_{ab}$ is reflection-symmetric, one
can always choose a flat metric $\eta$ so that $\sigma$ is also
reflection symmetric. Thus, to eliminate logarithmic translations
one only needs to ask that ${}^3\!E_{ab}$ be reflection symmetric
and then restrict oneself to asymptotic expansions (\ref{g1}) in
which $\sigma$ is also reflection symmetric. Finally, the symmetry
condition on ${}^3\!E_{ab}$ is not excessively restrictive. For
example, it is satisfied if $g$ is asymptotically a finite
superposition of Schwarzschild metrics with (possibly) distinct
asymptotic time translations \cite{ap}.

Next, let us consider super-translations. We first note a result due
to Beig and Schmidt \cite{bs}: Given a metric $g$ that admits an
expansion (\ref{g1}), one can always find another Minkowski metric
(which, for simplicity, we will denote by $\eta$ again) such that
the off-diagonal terms vanish to leading order. Thus, in the $(\rho,
\Phi)$ chart of the new Minkowski metric, we have:
\be \label{g2}  \dd s^{2}=(1+\f{2\sigma} {\rho}) \dd\rho^{2} \,+\,
\rho^2 (h_{ij} + \f{{}^1\!h_{ij}}{\rho})\, \dd\Phi^i \dd\Phi^j \,+\,
o(\rho^{-1})\, . \ee
where, by our assumption above, $\sigma(\Phi)$ is
reflection-symmetric on $\H$. For this metric $g$ one can calculate
the asymptotic Weyl curvature which can be decomposed into an
electric and a magnetic parts. The leading order magnetic part is a
tensor field tangential to $\H$, given by:
\be {}^3\!B_{ab} = \lim_{\rho\rightarrow \infty} {}^\star C_{acbd}
\h{x}^c\h{x}^d \,. \ee
${}^3\!B_{ab}$ admits a natural tensor potential $S_{ab} :=
{}^1h_{ab} + 2\sigma\, h_{ab}$ on $\H$: \, $B_{ab} =
\epsilon_{a}{}^{mn} D_m S_{nb}$ where $\epsilon$ and $D$ are the
alternating tensor and the derivative operator on the hyperboloid
$\H$ compatible with the metric $h_{ab}$, and indices are raised and
lowered by $h_{ab}$. Under super-translations, $x^a \rightarrow
\b{x}^a = x^a + S^a(\Phi)$, the field ${}^3\!B_{ab}$ itself is left
invariant while the potential transforms non-trivially:
\be\label{S}  \b\sigma = \sigma, \quad {}^1\!{\b{h}}_{ab} =
{}^1\!h_{ab} - 2 D_aD_b f -2f h_{ab}, \quad {\hbox{\rm so
that}}\quad \b{S}_{ab} = S_{ab} - 2 D_aD_b f -2f h_{ab},\ee
where the function $f$ on $\H$ is given by $f = f^a\h{x}_a$.
Finally, one can show that \emph{if} ${}^3\!B_{ab}$ vanishes, then
this transformation property of $S_{ab}$ is sufficient to ensure
that (via a suitable super-translation) we can always choose $\eta$
such that $S_{ab} =0$, i.e., ${}^1\!h_{ab} = -2\sigma h_{ab}$
\cite{ah,aa-ein,bs,ar}. The transformation property (\ref{S})
immediately implies that this exhausts the super-translation
freedom. The condition ${}^3\!B_{ab} =0$ is not excessively
stringent. For example, it is automatically satisfied if $g$ is
stationary \emph{or} axi-symmetric \cite{am}

Let us summarize. If the asymptotic Weyl curvature of a weakly
asymptotically flat space-time $(\M, g_{ab})$ is such that its
leading order term is \emph{purely electric} (i.e. if
${}^3\!B_{ab} =0$) and \emph{reflection symmetric} on $\H$, then
one can eliminate both super-translations and logarithmic
translations by strengthening the boundary conditions in a natural
fashion by requiring in (\ref{g2}) that $\sigma$ be reflection
symmetric and ${}^1\!h_{ab} = - 2\sigma h_{ab}$. Thus, the metrics
$g$ satisfying this additional conditions can be expanded as
\be \label{g3} \dd s^{2}=(1+\f{2\sigma} {\rho})\, \dd\rho^{2} \,+\,
(1-\f{2\sigma} {\rho})\, \rho^2\, h_{ij} \,\dd\Phi^i \dd\Phi^j \,+\,
o(\rho^{-1})\, , \ee
where $\sigma$ is reflection symmetric. Such space-times will be
said to be \emph{asymptotically flat at spatial infinity}. If a
given physical metric $g$ satisfies this condition with respect to
two distinct Minkowski metrics $\eta$ and $\b\eta$, then their
isometry groups agree to leading order: If $K^a$ is a Killing
vector of $\eta$, there exists a Killing vector $\b{K}^a$ of
$\bar\eta$ such that $\lim_{\rho\rightarrow \infty}\, K^a -
\b{K}^a = 0$. Hamiltonians generating these asymptotic symmetries
provide unambiguous definitions of conserved quantities at spatial
infinity. We will see that the additional conditions on $\sigma$
and ${}^1\!h_{ab}$ are essential for surface integral expressions
of the Lorentz angular momentum to be well-defined.

Thus, although the asymptotic structure is quite subtle in 4
dimensions, one can strengthen the `obvious' boundary conditions to
eliminate the logarithmic translations and super-translations and
yet admit a very large class of physically interesting examples.
(For a treatment of these issues in a canonical framework, see
\cite{bo}.)

\section{Action and the Covariant Phase space}
\label{s3}

We can now construct the Lagrangian and Hamiltonian descriptions
\emph{in the first order framework}. Our basic gravitational
variables will be co-triads $e_a^I$ and Lorentz connections
$A_a^{IJ}$ on space-time $\M$. Co-tetrads $e$ are `square-roots'
of metrics and the transition from metrics to tetrads is motivated
by the fact that tetrads are essential if one is to introduce
spinorial matter. $e_a^I$ is an isomorphism between the tangent
space $T_p(\M)$ at any point $p$ and a fixed internal vector space
$V$ equipped with a metric $\eta_{IJ}$ with Lorentzian signature.
The internal indices can be freely lowered and raised using this
fiducial $\eta_{IJ}$ and its inverse $\eta^{IJ}$. Each co-tetrad
defines a space-time metric by $g_{ab}:= e_a^I e_b^J \eta_{IJ}$
which also has signature $(-+++)$. Then the co-triad $e$ is
automatically orthonormal with respect to $g$. Since the
connection 1-forms $A$ take values in the Lorentz Lie algebra,
$A_a^{IJ} = -A_a^{JI}$. The connection acts only on internal
indices and defines a derivative operator
\[
D_a k_I :=\partial_a k_I + {A_{aI}}^J k_J \, ,
\]
where $\partial$ is a fiducial derivative operator which, as
usual, will be chosen to be flat and torsion free. As fundamental
fields, $e$ and $A$ are independent. However, the equation of
motion of $A$ implies that $A$ is compatible with $e$, i.e., is
fully determined by $e$. Therefore, boundary conditions on $A$ are
motivated by those on $e$. These in turn are dictated by our
discussion of asymptotics in section \ref{s2}.

In the Lagrangian and Hamiltonian frameworks we have to first
introduce the precise space of dynamical fields of interest. Let us
fix, once and for all, a co-frame ${}^o\!e_a^I$ such that $g^o_{ab}
= \eta_{IJ}\, {}^o\!e_a^I\, {}^o\!e_b^J$ is flat. The derivative
operator defined by ${}^o\!e_a^I$ will be denoted by $\partial_a$;
\, $\partial_a \, {}^o\!e_b^I =0$. The cartesian coordinates $x^a$
of $g^o_{ab}$ and the associated radial-hyperboloid coordinates
$(\rho, \Phi^i)$ will be used in asymptotic expansions. Discussion
of section \ref{s2} suggests that the co-triads $e_a^I$ should admit
an asymptotic expansion of order $1$. This suffices to obtain
well-defined 4-momentum. However, detailed analysis shows that to
define the Lorentz angular momentum one needs $e_a^I$ to admit an
expansion to order $2$ (see Sec. \ref{s4}). Therefore, we will
assume that $e_a^I$ can be expanded as:

\be\label{e} e = {}^o\!{e}(\Phi) + \f{{}^{1}e (\Phi)}{\rho} +
\f{{}^{2}e(\Phi)}{\rho^{2}} + o(\rho^{-2}) \ee
where ${}^1e_a^I$ is given by
\be \label{e1} {}^{1}{e_{a}^{I}} = \sigma(\Phi)\, (2 \rho_{a}
\rho^{I} -{}^{o}\!{e_{a}^{I}}) \ee
with a reflection symmetric $\sigma(\Phi)$ (see (\ref{g3}). Here and
in what follows
\be \rho_a = \partial_a \rho \quad {\rm and}\quad  \rho^I =
\eta^{IJ}\, \ez^a_J\, \rho_a \, .\ee
To appropriate leading orders, $A_a^{IJ}$ can be required to be
compatible with $e_a^I$ on the time-like world-tube
$\tau_{\infty}$ at spatial infinity which is part of the boundary
$\p\M$ of $\M$. This leads us to require that $A_a^{IJ}$ is of
asymptotic order $3$,
\be\label{A} A = {}^{o}\!{A}(\Phi) + \f{{}^{1}\!{A}(\Phi)}{\rho} +
\f{{}^{2}\!{A}(\phi)}{\rho^{2}} + \f{{}^{3}\!{A}(\Phi)}{\rho^{3}} +
o(\rho^{-3}) \ee
Compatibility of $A$ with $e$ and flatness of $\ez$ enables us to
set ${}^{0}\!{A} = {}^{1}\!{A} = 0$ and express ${}^2\!A$ as

\ba \label{A2} {}^{2}\!{A_{a}^{IJ}}(\Phi) &=& 2\rho^2\,
\partial^{[J}\,\left(\rho^{-1}\,\, {}^{1}\!e_{a}^{I]} \right)
\nonumber\\
%
%
&=& {2}{\rho}\, \left( 2\rho^{[I} \rho_a\,
\partial^{J]} \sigma \, - \, {}^o\!e_a^{[I} \partial^{J]} \sigma \,
-\, \rho^{-1}\,\,{{}^o\! e_a^{[I} \rho^{J]} \sigma} \right) \ea

(In spite of the explicit factors of $\rho$ the right side is in
fact independent of $\rho$ because $\p_a\sigma \sim \rho^{-1}\times$
(angular derivatives of $\sigma$).) We will not need the
corresponding expression of ${}^3\!A$ in terms of $e$ and therefore
demand compatibility between $A$ and $e$ only via (\ref{A2}).

\subsection{Action Principle}
\label{s3.1}

Consider as before the 4-manifold $\M$ bounded by space-like
surfaces $M_1$ and $M_2$. We will consider smooth histories $(e,A)$
on $M$ such that $(e,A)$ are asymptotically flat in the sense
specified above, and are such that $M_1,M_2$  are Cauchy surfaces
with respect to the space-time metrics $g$ defined by $e$, and the
pull-back of $A$ to $M_1,M_2$ is determined by the pull-back of $e$.
The last condition is motivated by the fact that, since the
compatibility between $e$ and $A$ is an equation of motion, boundary
values where this compatibility is violated are not of interest to
the variational principle. Finally it is convenient to partially fix
the internal gauge on the boundaries. We will fix a constant,
time-like internal vector $n^I$ so that $\p_a n^I =0$ and require
that the histories be such that $n^a := n^I e^a_I$ is the unit
normal to $M_1$ and $M_2$.

The first order gravitational action on these histories is given by
(see e.g. \cite{aa-book})
\begin{equation}\label{action}
S (e, A ) =
    -\,\frac{1}{2\k} \int_{\M} \Sigma^{IJ}\wedge F_{IJ}
   \, +\,\frac{1}{2\k} \int_{\p\M} \Sigma^{IJ} \wedge A_{IJ}
     \, ,
\end{equation}
where the 2-forms $\Sigma^{IJ}$ are constructed from the co-tetrads
and $F$ is the curvature $A$:
\[
\Sigma_{IJ}:= \textstyle{\f{1}{2}}\,\epsilon_{IJKL} e^K \wedge e^L
\quad {\rm and} \quad F_{I}{}^{J} = dA_{I}{}^{J} + A_{I}{}^{K}
\wedge A_{K}{}^{J} \qquad
 \, .
\]
As in more familiar field theories, the action now depends only on
the fundamental fields and their first derivatives. Although the
connection $A$ itself appears in the surface term at infinity,
action is in fact gauge invariant. Indeed, it is not difficult to
show that the compatibility between the pull-backs to $M_1$ and
$M_2$ of $e$ and $A$ and the property $\partial_a n^I=0$ implies
that, on boundaries $M_1$ and $M_2$, $\Sigma^{IJ} \wedge A_{IJ} =
2 K \epsilon_{abc}$ where $K$ is the trace of the extrinsic
curvature of $M_1$ or $M_2$ (see e.g., section 2.3.1 of
\cite{alrev}). Thus, on $M_1$ and $M_2$, the surface term in
(\ref{action}) is \emph{precisely the Gibbons-Hawking surface
term} with $C=0$ in (\ref{gh}). Therefore, these surface
contributions are clearly gauge invariant. This leaves us with
just the surface term at the time-like cylinder $\tau_{\infty}$ at
infinity. However, since $e$ has to tend to the fixed co-tetrad
$\ez$ at infinity, permissible gauge transformations must tend to
identity on $\tau_{\infty}$. Since the surface integral on
$\tau_{\infty}$ involves only the pull-back of $A$ to
$\tau_{\infty}$, it follows immediately that this surface integral
is also gauge invariant. Note incidentally that on $\tau_\infty$
this term is \emph{not} equal to the Gibbons-Hawking surface term
(because $\partial_a \rho^I$ falls off only as $1/\rho$).
Therefore, even if we were to assume compatibility between $e$ and
$A$ everywhere and pass to a second order action, (\ref{action})
would not reduce to the Gibbons-Hawking action with $C=0$. It is
also inequivalent to the Gibbons-Hawking prescription of setting
$C= K_o$ because (\ref{action}) is well-defined without reference
to any embedding in flat space.

Our boundary conditions allow us to rewrite this action as
\be S(e,A) = \f{1}{2 \kappa} \int_{M} d\Sigma \wedge A - \Sigma
\wedge A\wedge A \ee
Boundary conditions also imply that the integrand falls off as
$\rho^{-4}$. Since the volume element on any Cauchy slice goes as
$\rho^2\sin\theta \, \dd\rho \dd\theta \dd\phi$, the action is
manifestly finite \emph{even off shell} if the two Cauchy surfaces
$M_1$, $M_2$ are asymptotically time-translated with respect to
each other. Such space-times $\M$ are referred to as
\emph{cylindrical slabs}. In this sub-section our
discussion will be restricted to such space-times.%
\footnote{If the two Cauchy surfaces bounding $\M$ are
asymptotically boosted with respect to one another, $\M$ is called
a \emph{boosted slab}. For a boosted slab there is no guarantee
that the action would be finite \emph{off shell}. (The situation
is the same in Yang-Mills theory in Minkowski space.) However,
when equations of motion are satisfied the action reduces just to
the surface term which, because of our boundary conditions, is
proportional to $\textstyle{\int_{\tau_\infty}} \sigma d^3V$. The
asymptotic behavior \emph{in time} of $\sigma$ \cite{ap} implies
that this integral is finite. Thus the on shell result of
\cite{mm,mmv} is recovered in this first order framework.}

On the class of histories considered, it is easy to check that the
functional derivatives of the action are well defined with respect
to both $e$ and $A$. Variation with respect to the connection
yields $D \Sigma =0.$ This condition implies that the connection
$D$ defined by $A$ acts on internal indices in the same way as the
unique torsion-free connection $\nabla$ compatible with the
co-tetrad (which satisfies $\nabla_a\, e_b^I =0$). When this
equation of motion is satisfied, the curvature $F$ is related to
the Riemann curvature $R$ of $\nabla$ by
\[{F_{ab}}^{IJ}= {R_{ab}}^{cd}e_c^I e_d^J\, . \]
Varying the action with respect to $e_a^I$ and taking into account
the above relation between curvatures, one obtains Einstein's
equations $G_{ab} = 0$. Inclusion of matter is straightforward
because the standard matter actions contain only first derivatives
of fundamental fields without any surface terms and the standard
fall-off conditions on matter fields imply that the matter action
is finite on cylindrical slabs also off shell.

\subsection{Covariant Phase Space}
\label{s3.2}

We will now let $\M$ be $\R^4$. The covariant phase space $\Gamma$
will consist of smooth, asymptotically flat \emph{solutions $(e,A)$
to field equations} on $\M$. Thus, in contrast to section
\ref{s3.1}, $\M$ is not restricted to be a cylindrical slab nor are
the pull-backs of $(e,A)$ fixed on any Cauchy surfaces. Our task is
to use the action (\ref{action}) to define the symplectic structure
$\Omega$ on this $\Gamma$.

Following the standard procedure (see, e.g. \cite{abr}), let us
perform second variations of the action to associate with each phase
space point $(e,A)$ and tangent vectors $\delta_1 \equiv (\delta_1
e, \delta_1 A)$ and $\delta_2 \equiv (\delta_2 e, \delta_2 A)$ at
that point, a 3-form $J$ on $\M$, called the symplectic current:
\begin{equation}\label{symcur}
J(\gamma; \delta_1,\delta_2)=
   -\, \frac{1}{2\k}\,\,[ \delta_1 \Sigma^{IJ}\wedge \delta_2 A_{IJ}
        -\delta_2 \Sigma^{IJ} \wedge \delta_1 A_{IJ} ].
\end{equation}
Using the fact that the fields $(e,A)$ satisfy the field equations
and the tangent vectors $\delta_1, \delta_2$ satisfy the linearized
equations off $(e,A)$, one can directly verify that $J(\gamma;
\delta_1, \delta_2)$ is closed as guaranteed by the general
procedure involving second variations. Let us now consider a portion
$\wt{\M}$ of $\M$ bounded by two Cauchy surfaces $M_1, M_2 $. These
are allowed to be general Cauchy surfaces so $\wt{\M}$ may in
particular be a cylindrical or a boosted slab in the sense of
section \ref{s3.1}. Consider now a region $\t{\mathcal{R}}$ within
$\wt\M$, bounded by compact portions $\t{M}_1, \t{M}_2$ of $M_1$ and
$M_2$ and a time-like cylinder $\tau$ joining $\p \t{M}_1$ and $\p
\t{M_2}$. Since $dJ=0$, integrating it over $\t{\mathcal{R}}$ one
obtains
\be \int_{\t{M}_1} J + \int_{\t{M}_2} J + \int_{\tau} J =0 \ee
The idea is to take the limit as $\tau$ expands to the cylinder
$\tau_\infty$ at infinity. Suppose the first two integrals continue
to exist in this limit \emph{and} the third integral goes to zero.
Then, in the limit the sum of the first two terms would vanish and,
taking into account orientation signs, we would conclude that
$\int_M J$ is a 2-form on $\Gamma$ which is \emph{independent} of
the choice of the Cauchy surface $M$. This would be the desired
pre-symplectic structure. However, the issue of whether the boundary
conditions ensure that the integrals over Cauchy surfaces converge
and the flux across $\tau_\infty$ vanishes is somewhat delicate and
often overlooked in the literature.%
\footnote{Furthermore, even when such issues are discussed, one
often considers only the restricted action $\Omega(\delta,
\delta_V)$ of the pre-symplectic structure $\Omega$, where one of
the tangent vectors, $\delta_V$, is associated with an asymptotic
symmetry $V^a$ on $\M$ in the sense discussed in section \ref{s4}
because, as we will see, it is this restricted action that directly
enters the discussion of conserved quantities. Typically the 3-forms
$\Omega(\delta_1, \delta_V)$ on $\M$ have a better asymptotic
behavior that the generic 3-forms $\Omega(\delta_1, \delta_2)$.
However, unless $\Omega(\delta_1,\delta_2)$ is well-defined for all
$\delta_1, \delta_2$, one does not have a coherent Hamiltonian
framework and cannot start constructing conserved quantities.}
If either of these properties failed, we would not obtain a
well-defined symplectic structure on $\Gamma$.

Let us first consider the integral over the time-like boundary
$\tau$. As $\tau$ tends to $\tau_\infty$, the integrand $J_{abc}
\epsilon^{abc}$ tends to
\be \label{flux} \lim_{\tau \rightarrow \tau_\infty}\,\,
\left(\delta_{[1}\f{{}^{1}\!{\Sigma}_{ab}}{\rho}\right) \,\wedge\,
\left(\delta_{2]} \f{{}^{2}\!{A_c}}{\rho^{2}}\right)\,
\epsilon^{abc}\, =\,  \lim_{\tau \rightarrow \tau_\infty}\,\,
\epsilon_{IJKL}\,\, {}^{o}\!{e_{a}^{K}}\, (\delta_{[1}
{}^{1}\!{e_{b}^{L}})\, (\delta_{2]} {}^{2}\!{A_{c}^{IJ}})\,
\rho^{-3}\,\,\epsilon^{abc}\, \ee
where $\epsilon^{abc}$ is the metric compatible 3-form on $\tau$.
Since the volume element on $\tau$ goes as $\rho^3$, the integral of
the symplectic flux over $\tau$ has a well-defined limit. But the
key question is if the limit is \emph{zero}. If not, there would be
a leakage of the current $J$ at spatial infinity and the symplectic
structure would not be well-defined. Let us evaluate this term using
the expressions (\ref{e1}) and (\ref{A2}) of ${}^1\!e$ and
${}^2\!A$. Then the second term in (\ref{flux}) reduces to
\be \label{flux2} \epsilon_{IJKL}\,{}^o\!e_a^K\, \left(2 \rho_b
\rho^L - {}^o\!e_b^L\right) \delta_1 \sigma\,\, \left(2 \rho^I
\rho_c \partial^J \delta_2 \sigma - {}^o\!e_c^I \partial^J \delta_2
\sigma - {}^o\!e_c^I \rho^J \delta_2 \sigma\right)\,\,\rho^{-3}
\epsilon^{abc} \quad - \,\,1\leftrightarrow 2 \ee
The term containing $\delta_1\sigma \delta_2\sigma$ vanishes because
of anti-symmetrization while the remaining terms containing
derivatives of $\delta_2 \sigma$ vanish because the normal $\rho^c$
to $\tau$ is contracted either with $\epsilon^{abc}$ or the
derivative of $\sigma$. Thus, our boundary conditions imply that the
symplectic flux across $\tau_\infty$ vanishes.

The next question is whether the integral over $\t{M}_1$ (and
$\t{M}_2$) continues to be well-defined in the limit as we
approach $M_1$ (resp. $M_2$). The leading term is again given by
the integral of (\ref{flux}) over $M_1$, the only difference being
that $\epsilon^{abc}$ is now the metric compatible 3-form on
$M_1$. Since the volume element on $M_1$ goes as $\rho^2\, d\rho
d^2\Phi$, power counting argument says that the integral of this
leading term can be logarithmically divergent. However, one can
again expand out the leading term as (\ref{flux2}) and show that
it in fact vanishes. Since the remaining integrand falls off at
least as fast as $1/\rho^4$, the integral over $M$ converges.
Thus, because of our boundary conditions, we are led to a
well-defined pre-symplectic structure, i.e., a closed 2-form, on
$\Gamma$
\be \Omega(\delta_{1},\delta_{2})= \f{1}{2 \kappa}\, \int_{M} {\rm
Tr}\, \left[ \delta_{1} \Sigma \wedge \delta_{2} A - \delta_{2}
\Sigma \wedge \delta_{1} A \right]\, ,\ee
where $M$ is any Cauchy surface in $\M$ and trace is taken over the
internal indices. $\Omega$ is not a symplectic structure because it
is degenerate. The vectors in its kernel represent infinitesimal
`gauge transformations'.  The physical phase space is obtained by
quotienting $\Gamma$ by gauge transformations and inherits a true
symplectic structure from $\Omega$. We will not carry out the
quotient however because the calculation of Hamiltonians can be
carried out directly on $(\Gamma, \Omega)$.

\section{Generators of asymptotic Poincar\'e Symmetries}
\label{s4}

Let $V^a$ be a vector field on $\M$ representing an asymptotic
Poincar\'e symmetry, a Killing vector field of one of the flat
metrics $\eta_{ab}$ in $\Gamma$. Then at any point $(e,A)$ of
$\Gamma$,\, the pair $(\L_V e, \L_V A)$ of fields satisfies the
linearized field equations, whence $\delta_V := (\L_V e, \L_V A)$ is
a vector field on $\Gamma$. (In the definition of the
Lie-derivative, internal indices are treated as scalars; thus $\L_V
e_a^I = V^b\partial_b e_a^I + e_b^I
\partial_a V^b$.) The question is whether $\delta_V$ is a phase
space symmetry, i.e., whether it satisfies $\L_{\delta_V} \Omega
=0$.

Consider the 1-form $X_V$ on $\Gamma$ defined by
\be\label{defx} X_V (\delta)= \Omega(\delta, \delta_{V}). \ee
$\L_{\delta_{V}}\, \Omega = 0$ on $\Gamma$ if and only if $X_V$ is
closed, i.e.,
\[\D X_V =0\]
where $\D$ denotes the exterior derivative on (the infinite
dimensional) phase space $\Gamma$. If this is the case then, up to
an additive constant, the Hamiltonian is given by
\[\D H_V = X_V.\]
The constant is determined by requiring that all Hamiltonians
generating asymptotic symmetries at the phase space point $(\ez,
A=0)$ corresponding to Minkowski space-time must vanish. To
calculate the right side of (\ref{defx}), it is useful to note the
Cartan identities
\be\label{liet}
    \L_V A = V \cdot F + D(V\cdot A) \quad{\rm and} \quad
    \L_V \Sigma = V\cdot D\Sigma + D(V\cdot \Sigma) -[(V\cdot A),\Sigma]
\ee
Using these, the field equations satisfied by $(e,A)$ and the
linearized field equations for $\delta$, one obtains the required
expression of $X_V$:
\be\label{xt} X_V(\delta) := \Omega(\delta,\delta_V)=
-\,\frac{1}{2\k} \oint_{S_\infty} {\rm Tr}\left[(V\cdot A) \delta
\Sigma - (V\cdot \Sigma) \wedge\delta A \right] \, . \ee
Note that the expression involves integrals \textit{only} over the
2-sphere boundary $S_\infty$ of the Cauchy surface $M$ (i.e., the
intersection of $M$ with the hyperboloid $\H$ at infinity); there
is no volume term. This is a reflection of the fact that the
theory is diffeomorphism invariant.

\subsection{Energy-Momentum}
\label{s4.1}
Let us begin by setting $V^a=T^a$, an infinitesimal asymptotic
translation. Since $\delta \Sigma \sim 1/\rho,\,\, A \sim 1/\rho^2$
and since the area element of the 2-sphere $S_\infty$ at infinity
grows as $\rho^2$, the first term on the right side of (\ref{xt})
vanishes in the limit and we are left with
\be X_T(\delta) := \Omega(\delta,\delta_{T}) = \f{1}{2 \kappa}
\oint_{S_{\infty}} {\rm Tr} [(T \cdot \Sigma) \wedge \delta A] \ee
which is manifestly well-defined. Inserting the asymptotic forms
of the connection (\ref{A2}) and tetrad (\ref{e}) we find:
\be X_{T}(\delta)  = \f{2}{\kappa}\, \oint_{S_\infty} [(\rho_a
T^a)\, n^{b}\partial_{b}\delta \sigma \,+\, \delta \sigma\, (n_a
T^a)] \dd^2S_o \ee
where $n_{b}$ is the unit normal to the 2-sphere $S_\infty$ within
the hyperboloid $\H$ and $d^2S_o$ is the area element of the
\emph{unit} 2-sphere. Since the only dynamical variable in the
integrand is $\sigma$, we can pull the $\delta$ out of the
integral and obtain the Hamiltonian $H_T$ generating the
asymptotic translation $T^a$:
\be \label{4P} H_{T} = \f{2}{\kappa} \oint_{S_\infty} [(\rho_a
T^a)\, n^{b}\partial_{b} \sigma \,-\, \sigma (n_a T^a)] \dd^2S_o \ee
Had we selected a translational Killing field $\bar{T}^a$ of another
flat metric $\bar{\eta}_{ab}$ in our phase space $\Gamma$, we would
have obtained the same answer because $\bar{T}^a - T^a =
o(\rho^{-1})$.

Taking our transformation to be a unit time-translation which is
asymptotically orthogonal to the Cauchy surface $M$ under
consideration (and hence to $S_\infty$) we find the energy to be:
\be\label{E}  E = \f{2}{\kappa} \oint_{S_\infty} \sigma\, \dd^2S_o
\ee
Similarly, if $T^a$ is a space-translation which is asymptotically
tangential to $M$, we find
\be \label{3P} \vec{P}\cdot \vec{T} = \f{2}{\kappa}\,
\oint_{S_\infty} (\rho_a T^a)\, (n^bD_b \sigma)\, \dd^2S_o\ee
where $D$ is the derivative operator on the unit hyperboloid $(\H,
h_{ab})$. Note that $\rho_a T^a$ are the `$\ell =1$' spherical
harmonics on $S_\infty$ determined by translations $T^a$.

Thus the energy momentum is determined directly by $\sigma$. It
follows from \cite{ah,aa-ein,bs} that $\sigma$ satisfies the
hyperbolic equation $D^aD_a \sigma + 3\sigma =0$ on $(\H, h_{ab})$.
Thus, its initial data consists of the pair $(\sigma, \dot\sigma=
n^a D_a \sigma)$ on a 2-sphere cross-section of $\H$. Energy is
given by the `$Y_{00}$' component of the first piece of this data
while the momentum by the `$Y_{1m}$' components of the second piece.
Finally, recall that the reflection through the origin of Minkowski
space induces an isometry on the unit hyperboloid. Therefore
\emph{every} solution $\sigma$ to the hyperbolic equation can be
decomposed into a part $\sigma_{(E)}$ which is even under this
reflection and a part $\sigma_{(O)}$ which is odd. Each satisfies
the hyperbolic equation separately. Our boundary conditions require
that $\sigma$ be even. However it is easy to verify that, even if
this condition had not been imposed, only the even part
$\sigma_{(E)}$ contributes non-trivially to the expressions
(\ref{E}) and (\ref{3P}) of energy and momentum.

\subsection{Relativistic Angular Momentum}
\label{s4.2}

Let us now set $V^a= L^a$, an infinitesimal asymptotic Lorentz
symmetry. For definiteness, we will assume that it is a Lorentz
Killing field of $\eta_{ab}:= \eta_{IJ}\, \ez^I_a \ez^J_b$ so that
it is tangential to the $\rho = {\rm const}$ hyperboloids $\H$.
The question is whether the vector field $\delta_L$ on $\Gamma$ is
Hamiltonian. Let us begin by examining the 1-form $X_L$ on $\Gamma$.
Using (\ref{xt}), we have:

\be \label {xL} X_{L}(\delta) := \Omega(\delta, \delta_L) = \f{1}{2
\kappa} \lim_{\rho \rightarrow \infty} \oint_{S_{\rho}}\, {\rm Tr}\,
[(L \cdot A)\, \delta \Sigma\, +\, (L \cdot \Sigma) \wedge \delta A
]\ee
where $S_\rho$ is the 2-sphere intersection of the $\rho = {\rm
constant}$ hyperboloid $\H_\rho$ with the Cauchy surface $M$ used to
evaluate the symplectic structure. Now, as $\rho$ tends to infinity,
$\Sigma$ has a well-defined, non-zero limit, $A \sim \rho^{-2},\,
\delta A \sim \rho^{-2},\, \delta \Sigma \sim \rho^{-1}$ and $L \sim
\rho$. Therefore the second term in (\ref{xL}) is potentially
divergent. Using the asymptotic form (\ref{A2}) of $\delta A$, it
follows that the second term is proportional to
\be \lim_{\rho \rightarrow \infty}\, \oint_{S_\rho} (\delta \sigma\,
L^a n_a)\,\, \rho^{-2} \dd^2S\ee
where $n_a$ is the unit normal to $S_\rho$ within the hyperboloid
$\H_\rho$ (or equivalently, to the Cauchy surface $M$) and $\dd^2S$
is the volume element on $S_\rho$ (which grows as $\rho^2$). If we
were interested in the rotational sub-group of the Lorentz group
adapted to $M$, the vector field $L^a$ would be tangential to $M$,
whence this term would vanish.%
\footnote{One might first think that since the integral vanishes for
all rotations, by changing the Lorentz frame defining the rotation
subgroup one would be able to show that the integral vanishes also
for boosts. This turns out not to be correct. To handle Lorentz
boosts one needs a genuinely stronger asymptotic condition. As
discussed below this is provided by the reflection symmetry of
$\sigma$ discussed in section \ref{s2}. This point is often not
realized because much of the literature focuses only on rotational
subgroups. See, e.g. \cite{mm,mmv}.}
But for a boost, $L^a$ is proportional to $n^a$ whence the potential
divergence survives. Recall, however, that our boundary conditions
require that $\sigma$ be invariant under reflection symmetry. For a
Lorentz boost, on the other hand, $L^an_a$ is proportional to
$Y_{1m}$, and therefore odd, whence the integral vanishes. Thus,
thanks to the parity condition on $\sigma$, the 1-form $X_L$ is
well-defined on $\Gamma$.

To extract the Hamiltonian from $X_L$, we need to pull $\delta$
out of the integral. This is possible because the reflection
symmetry again implies that the potentially divergent contribution
from ${}^2A_a^{IJ}$ vanishes. Furthermore, using the asymptotic
forms (\ref{e}) and (\ref{A2}) it follows that contributions from
$(L \cdot {}^{1}{\Sigma})\, \wedge\, {\delta\, {}^{2}\! A}$ and
$(L \cdot {}^{2}\!{A})\,\wedge\, {\delta\, {}^{1}\Sigma}$ cancel
each other. Consequently, $X_L = \D H_L$ where the Hamiltonian
generating the Lorentz transformation $L^a$ is given by
\be \label{J} H_{L}= - \f{1}{2\kappa}\,\oint_{S_\infty} {\rm
Tr}\,(\h{L} \cdot {}^{o}{\Sigma})\, \wedge {}^{3}\!{A}\, , \ee
where $\h{L}^a = L^a/\rho$ is the Lorentz Killing field on the
\emph{unit} hyperboloid $(\H, h_{ab})$. Note that in contrast to the
energy momentum, the angular momentum is not determined by the
leading order deviation of $(e,A)$ from the ground state $(\ez,
A=0)$ but by sub-leading terms.

\section{Relation to the Spi framework}
\label{s5}

The boundary conditions we imposed in section \ref{s3} to construct
the Lagrangian and Hamiltonian frameworks imply that the space-time
admits a conformal completion with conformal factor $\omega =
\rho^{-2}$ in which spatial infinity is represented by a single
point $i^o$. The conformally rescaled metric $\h{g}_{ab} = \omega^2
g_{ab}$ can be shown to have the regularity needed in the so-called
Spi-framework%
\footnote{Spi stands for spatial infinity and rhymes with scri
that represents null infinity.}
\cite{ah, aa-ein}. These conditions in turn imply that various
physical fields admit a direction dependent limit as one
approaches $i^o$ in space-like directions and can therefore be
regarded as smooth fields on the unit hyperboloid $\H_o$ in the
tangent space at $i^o$. The boundary conditions we imposed have
been shown to eliminate the logarithmic translations \cite{aa-log}
and supertranslations \cite{ah,aa-ein} also in the Spi-framework,
reducing the asymptotic group to the Poincar\'e group. Using field
equations in the physical space-time, fields on $\H_o$ were shown
to satisfy certain equations and these were used to define
Poincar\'e momentum in terms of the asymptotic Weyl curvature. The
asymptotic field equations made it evident that these quantities
are conserved, i.e., are independent of the choice of the 2-sphere
cross-section of $\H_o$ used in their evaluation.

In section \ref{s4}, by contrast, we were led to the expressions
of Poincar\'e momenta using Hamiltonian considerations and our
final expressions are surface integrals involving asymptotic forms
of triads and connections rather than the Weyl curvature. It is
natural compare the underlying assumptions and ask for the
relation between these quantities and those obtained in the Spi
framework.

\subsection{Energy-momentum}
\label{s5.1}

In the Spi framework, the total energy momentum $P_a$ is a 4-vector
in the tangent space at $i^o$. Let $T^a$ be an asymptotic
translation in the physical space-time. Then it defines a vector
$T^a_o$ at $i^o$ and corresponding component $T^a_oP_a$ of the
4-momentum is given by \cite{ar,ah,aa-ein}
\be T^a_o p_a \,=\, -\,\f{1}{\k} \oint_{S} {\bf E}_{ab}\,
\h{T}^b_o\, n^a d^2S \, .\ee
Here, ${\bf E}_{ab}$ is the `electric part' of the asymptotic Weyl
curvature, $\h{T}^a_o = h^a_b T^b$ is the conformal Killing field
on the unit hyperboloid $(\H_o, h_{ab})$ in the tangent space
$T_{i^o}$, and $S$ any 2-sphere cross section of $\H_o$. In terms
of the physical space-time fields used in this paper, ${\bf
E}_{ab}$ is given by ${\bf E}_{ab} = {}^3\!E_{ab} \equiv
\lim_{\rho\rightarrow \infty} \rho^3 \, C_{ambn}\, \rho^m\rho^n$.
Thus, as mentioned in section \ref{s2}, ${}^3\!E_{ab}$ is the
leading order electric part of the asymptotic Weyl tensor, where
the electric and magnetic decomposition is carried out using a
$\rho = {\rm const}$ foliation.

To relate this $P_a$ to that defined in section \ref{s4}, let us
recall that $\sigma$ serves as a scalar potential of
${}^3\!E_{ab}$:
\be {}^3\!E_{ab} =  -\,\left(D_a D_b \sigma + \sigma h_{ab}\right)
\ee
Let us choose $S$ to be the intersection of $\H_o$ with a space-like
plane in $T_{i^o}$ and let $T^a$ be unit and orthogonal to this
plane. The question then is whether $T^a_o p_a$ equals the energy
(\ref{E}) defined in section \ref{s4} using Hamiltonian
considerations.

The trace-free property of  ${}^3\!E_{ab}$ implies that $\sigma$
satisfies the hyperbolic equation $D^aD_a \sigma + 3 \sigma =0$
and it is straightforward to verity that, \emph{on the specific
cross-section} $S$ of $\H_o$ we chose, $D_a \h{T}^b_o =0$.
Therefore on this $S$, we have: $E_{ab} \h{T}^a_o n^b =
E_{ab}\h{T}^a_o\h{T}^b_o  = -\h{T}^a_oD_a \h{T}^b_o D_b \sigma
+\sigma = -(\Delta+2) \sigma$, where $\Delta$ is the Laplacian on
$S$. Hence, we have
\be T^a P_a \, = \, \f{2}{\k}\, \oint_S\, \sigma d^2 S \ee
which agrees with the expression (\ref{E}) of energy obtained from
Hamiltonian considerations.

Next, let $T^a$ be a space-translation tangential to the 3-plane
whose intersection with $\H_o$ defined $S$. Then $\h{T}^a_o n^b
h_{ab} =0$ and ${D}_a V^a = 2 V^a\rho_a$ on $S$ where $\rho_a$ is
the unit normal to $\H_o$.  Using the fact that $D_a n^b$ vanishes
on $S$ we obtain,
\be T^a P_a \, =\, \f{2}{\k} \oint_S (\h{T}^a_o\rho_a)\, (n^bD_b
\sigma)\, d^2S \ee
which agrees with the expression (\ref{3P}) of the 3-momentum
obtained from Hamiltonian considerations.

Thus, the energy-momentum obtained using Hamiltonian considerations
agrees with that obtained using just the asymptotic field equations
in the $i^o$ framework. A detailed examination shows that in both
frameworks one can impose substantially weaker boundary conditions
to arrive at this expression of energy-momentum. In particular, one
only needs to require that $e$ be of asymptotic order just $1$
(rather than $2$) and, furthermore, one can drop the requirement
that the $1/\rho^3$-part of the asymptotic Weyl curvature be pure
electric and reflection symmetric, thereby allowing both logarithmic
translations and super-translations.

\subsection{Relativistic Angular Momentum}
\label{s5.2}

The situation with angular momentum is more subtle. To define
Lorentz angular-momentum one has to get rid of the
super-translation ambiguity also in the Spi framework. Indeed, the
procedure we followed in section \ref{s2} merely mimicked the Spi
strategy of requiring that the $1/\rho^3$ contribution to the
magnetic part $B_{ab}$ of the physical Weyl tensor should vanish
and then setting its tensor potential to zero. However, in
striking contrast to what we found in the discussion of
Hamiltonians $H_L$ in section \ref{s4.2}, the Spi angular momentum
is well-defined without having to require that $\sigma$ be
reflection symmetric. Indeed, the Spi expression is constructed
directly from ${}^4\!B_{ab}$ ---the $1/\rho^4$ part of $B_{ab}$---
and this field is insensitive to logarithmic translations
\cite{aa-log}. Furthermore, as we will now show the final
Spi-expression coincides with the expression (\ref{J}) of $H_L$ we
found in section \ref{s4.2}. Why then was the parity condition
essential in our derivation of (\ref{J})? It was necessary
because, in the Hamiltonian approach, not only should the
expression of angular momentum be well defined, but it should also
be the generator of asymptotic Lorentz transformations. More
precisely, the parity condition is needed to show that the 1-form
$X_L$ on the phase space $\Gamma$ is well-defined and exact, i.e.,
that the right side of (\ref{J}) has the interpretation of
Hamiltonians generating Lorentz rotations on the gravitational
phase space.

In the Spi framework angular momentum is constructed from the
$1/\rho^4$ part of the magnetic Weyl tensor
\be\label{beta} {}^4\!B_{ab} := \lim_{\rho \rightarrow
\infty}\rho^4\,\, {}^\star C_{ambn} \rho^m \rho^n \ee
where, as before, $\rho^m$ is the unit normal to the $\rho={\rm
const}$ hyperboloids. Every Lorentz Killing field $\h{L}^a$ on the
unit hyperboloid is of the form $\h{L}^a = F^{ab}_o \rho_b$ where
$F^{ab}_o$ is a constant skew tensor in Minkowski space $(\M, \eta)$
and therefore has a dual Lorentz Killing field defined by ${}^\star
\h{L}^a = {}^\star F_o^{ab} \rho_b$. The angular momentum
$J_{\h{L}}$ associated with the Lorentz Killing field $L^a$ is
defined as:
\be \label{J2} J_{\h{L}} = \oint_S\, {}^4\!B^{ab}\,\, {}^\star
\h{L}_a\, \epsilon_{bmn} dS^{mn}\ee
where as before $S$ is any 2-sphere cross-section of $\H_o$ and
$\epsilon$ is the volume 3-form on the unit hyperboloid $(\H_o,
h_{ab})$. To relate $J_{\h{L}}$ with the Hamiltonian $H_L$ of
(\ref{J}), we first write the Weyl tensor in (\ref{beta}) as
$C_{abcd} = F_{ab}{}^{IJ}\, e_{cI} e_{dJ}$ and use the fact that
$F_{ab}{}^{IJ}$ is given by $F_{ab}{}^{IJ} = 2\p_{[a}\, A_{b]} +
[A_a, A_b]^{IJ}$. Then, by using the asymptotic expansions (\ref{e})
and (\ref{A2}), one obtains%
\footnote{The calculation is significantly simplified by noting
that (\ref{e}) and (\ref{A}) imply that the $1/\rho^3$-part
${}^3\!B_{ab}$ of $B_{ab}$ vanishes, whence one can retain just
the $1/\rho^4$ terms in the expression of ${}^4\!B_{ab}$ in terms
of $A$.},
\ba J_{\h{L}}  &=& - \f{1}{2\kappa}\,\oint_{S} {\rm Tr}\,(\h{L}\cdot
{}^{o}{\Sigma})\, \wedge {}^{3}\!{A}\, ,\nonumber\\
&=& H_L.
 \ea
Thus the angular momentum constructed from asymptotic field
equations in the Spi framework agrees with that obtained in this
paper from Hamiltonian considerations. It therefore also follows
that in an asymptotically flat \emph{axi-symmetric} space-time
$H_L$ reproduces the Komar integral if $L^a$ is chosen to be the
rotational Killing field, and in an asymptotically flat
\emph{stationary} space-time, it yields the same angular momentum
dipole moment as that constructed from the stationary Killing
field \cite{am}.

\section{Discussion}
\label{s6}

In this paper we have shown that in the first order formalism based
on co-triads and Lorentz connections, the Lagrangian and Hamiltonian
frameworks can be constructed without having to introduce an
infinite counter term subtraction in the action. However, since in
four space-time dimensions physical metrics approach the flat metric
only as $1/\rho$ at spatial infinity, the obvious boundary
conditions allow one to make supertranslations and logarithmic
translations. The asymptotic symmetry group is then larger than the
Poincar\'e group. If one is interested only in energy momentum,
these ambiguities can be ignored because one can still single out a
well-defined 4-dimensional group of asymptotic translations. To have
a well-defined angular momentum, on the other hand, the obvious
boundary conditions used in much of the older literature are too
naive; they have to be carefully strengthened to reduce the
asymptotic symmetry group to the Poincar\'e group. When this is
done, the Hamiltonians generating asymptotic Poincar\'e
transformations provide us with expressions of energy-momentum and
Lorentz angular momentum. These agree with the expressions obtained
in the Spi framework based of asymptotic field equations
\cite{ah,aa-ein}. Therefore they also agree with conserved
quantities defined by other methods in restricted cases with exact
symmetries \cite{am}.

For simplicity, in this paper we focused on vacuum Einstein's
equations. However, inclusion of standard matter ---in particular,
scalar, Maxwell and Yang-Mills fields--- with standard boundary
conditions used in Minkowski space is straightforward. There are no
surface terms in the action associated with matter and the
expressions of the Hamiltonians generating asymptotic Poincar\'e
transformations are formally the same as the ones we found. In
particular, the Hamiltonians consist entirely of 2-sphere surface
integrals at spatial infinity and their integrands do not receive
any explicit contributions from matter. Matter makes its presence
felt through constraint equations which, in response to matter,
modify the asymptotic gravitational fields. Finally, in higher
dimensions, the asymptotic structure is considerably simpler because
the physical metric approaches the flat metric as $1/\rho^2$ or
faster. This issue is discussed in the accompanying paper \cite{as}.

We conclude with a comment. In non-commutative geometry, in place
of Riemannian geometry, one introduces a spectral triplet
$(\mathcal{A}, \mathcal{H}, D)$ consisting of a non-commutative
$C^\star$ algebra $\mathcal{A}$, a representation of it on a
Hilbert space $\mathcal{H}$ and a Dirac operator $D$ acting on
$\H$. A certain choice of the triplet is made to describe (a
generalization of) the standard model of particle physics together
with Einstein gravity. Rather general symmetry considerations then
lead to a so-called `spectral action' from which dynamics can be
derived \cite{cc1}. It has been known for some time that an
asymptotic expansion of this action can be performed to make
contact with the low energy physics and the first terms reproduce
the Einstein-Hilbert action with a cosmological constant. Recently
it was realized \cite{cc2} that the spectral action can be
naturally extended to incorporate the presence of boundaries and
the asymptotic expansion of the new action produces
\emph{precisely} the Einstein Hilbert action with the
Gibbons-Hawking counter term for $C=0$ (see (\ref{gh})). This is
an exciting development. However, as the discussion of section
\ref{s1} shows, in the asymptotically flat context this action has
severe limitations and the extension of the spectral action to
incorporate the boundary term \cite{cc2} was motivated using
precisely the Hamiltonian formulation in the asymptotically flat
context. More generally, the non-commutative framework has been
developed primarily for the Riemannian signature and passage to
the Lorentz signature is contemplated via a Wick transform in the
asymptotically flat context. Therefore asymptotic considerations
of \cite{mm} and this paper are directly relevant to the spectral
action approach. The natural question then is: Can the spectral
action framework be further generalized so that the leading terms
in the asymptotic expansion yields an action which is free from
the drawbacks of (\ref{gh})? The first order framework discussed
in this paper presents a natural avenue for such a generalization.
Indeed, the gravitational sector of the non-commutative geometry
requires a spin-bundle ---and hence a frame field $e$--- as well a
Dirac operator ---i.e., a spin connection $A$. However, in the
non-commutative framework the two are in essence compatible with
one another from the beginning. The question is whether one can
extend the framework so that they are independent to begin with
and made compatible only by equations of motion. The spectral
action in such a generalization could then descend to
(\ref{action}) upon a suitable asymptotic expansion. Quite apart
from this specific application, such a `first order' framework in
non-commutative geometry appears also to lead to mathematical
structures which are interesting in their own right, and could
provide a technical bridge between non commutative geometry and
loop quantum gravity.

\section*{Acknowledgment:} We would like to thank Don Marolf and
Nigel Higson for stimulating discussions. This work was supported
in part by the NSF grants PHY-0456913 and OISE-0601844, the
Alexander von Humboldt Foundation and the Eberly research funds of
Penn State.

\begin{appendix}
\end{appendix}

\end{document}